\documentclass[runinauthor,superscriptaddress,twocolumn]{revtex4}
\usepackage{natbib}
\usepackage[dvipdf]{graphicx}
\usepackage{amssymb}
\usepackage{amsmath}
\usepackage{booktabs}
\usepackage{lscape}
\usepackage{txfonts}
\usepackage{color}

\usepackage{ulem}  
\begin{document}
\title{Coherent dynamics of mixed Frenkel and Charge Transfer Excitons in Dinaphtho[2,3-b:2$^\prime$3$^\prime$-f]thieno[3,2-b]-thiophene Thin Films: 
The Importance of Hole Delocalization}
\author{Takatoshi Fujita}
\email{t.fujita@kuchem.kyoto-u.ac.jp}
\affiliation{Department of Chemistry, Graduate School of Science, Kyoto University, Kyoto 606-8502, Japan}

\author{Sule Atahan-Evrenk}
\affiliation{TOBB University of Economics and Technology, Sogutozu, Ankara 06560, Turkey}
\author{Nicolas P. D. Sawaya}
\affiliation{Department of Chemistry and Chemical Biology, Harvard University, Cambridge, Massachusetts 02138, USA}

\author{Al\'{a}n Aspuru-Guzik}
\email{aspuru@chemistry.harvard.edu}
\affiliation{Department of Chemistry and Chemical Biology, Harvard University, Cambridge, Massachusetts 02138, USA}

\begin{abstract}
Charge transfer states in organic semiconductors play crucial roles in processes such as singlet fission and exciton dissociation at donor/acceptor interfaces. 
Recently, a time-resolved spectroscopy study of dinaphtho[2,3-b:2$^\prime$3$^\prime$-f]thieno[3,2-b]-thiophene (DNTT) thin films provided evidence for the formation of mixed Frenkel and charge-transfer excitons after the photoexcitation. Here we investigate optical properties and excitation dynamics of the DNTT thin films by combining ab initio calculations and a stochastic Schr\"{o}dinger equation.
Our theory predicts that the low-energy Frenkel exciton band consists of 8 to 47\% CT character. The quantum dynamics simulations show coherent dynamics of Frenkel and CT states in 50 fs after the optical excitation. We demonstrate the role of charge delocalization and localization in the mixing of CT states with Frenkel excitons as well as the role of their decoherence.
\end{abstract}

\maketitle

Organic semiconductors (OSCs) are widely investigated as candidates for inexpensive and flexible materials for photovoltaics and other optoelectronic applications.~\cite{Bendikov2004,Anthony2006,Bredas2009}.  In recent years, new OSCs have been designed and improved through computational modeling~\cite{Blouin2008,Sokolov2011,Li2012} and virtual high-throughput screening~\cite{Hachmann2014,Boyle2011,Pyzer-Knapp2015}. Modeling energy and charge transport properties is also essential for understanding structure-property relationships and thus for rationally designing novel OSCs~\cite{Bredas2004,Wang2010,Troisi2011,Baumeier2012,Mei2013,Sule2014,Jiang2016}.

The low-energy optical excitations in OSCs lead to the formation of a bound electron--hole (e--h) pair, a Frenkel exciton. Excitonic properties of organic crystals are substantially different from those of isolated molecules, owing to excitonic couplings, near-field interactions between electronic transitions~\cite{Scholes2006,Spano2010,Kuhn2011,Saikin2013}. 
Interactions with a charge-transfer (CT) state---a state in which the electron and hole are located on spatially separated regions---can also affect the optical properties of an exciton, as has been demonstrated by several theoretical studies~\cite{Hoffmann2002,Gisslen2009,Yamagata2011,Fujimoto2013,Sharifzadeh2013,Sharifzadeh2015}.
Experimentally, the degree of CT character has been studied by momentum-dependent electron-loss spectroscopy~\cite{Schuster2007,Roth2012,Roth2013} or electroabsorption spectroscopy~\cite{Sebastian1981,Moller2000,Hass2010}.
CT states can act as precursors for interfacial CT states~\cite{Zhu2009,Clarke2010,Bakulin2012}; mixing of CT states with Frenkel excitons would facilitate charge separation in photovoltaic materials. 
They have also gained recent attention due to their relevance to singlet fission~\cite{Smith2010,Chan2013,Monahan2015}, in which singlet to triplet conversion can proceed via sequential CT steps~\cite{Beljonne2013,Berkelbach2013a,Berkelbach2013b}.

Here we focus on dinaphtho[2,3-b:2$^\prime$3$^\prime$-f]thieno[3,2-b]-thiophene (DNTT), a $p$-type OSC originally introduced by Takimiya and co-workers~\cite{Yamamoto2007}.
DNTT and its derivatives~\cite{Sule2010,Sokolov2011,Kang2011,Kang2012,Xie2013,Kinoshita2014,Ishino2014,Miyata2015} have gained attention due to their high hole mobility values and air stability. A recent time-resolved spectroscopy study by Ishino et al.~\cite{Ishino2014} concludes that mixed Frenkel and CT excitons are formed after the optical excitation. Although the degree of CT character in excited states has been reported as described in the earlier paragraph, its role in excitation dynamics has remained unclear. Furthermore, there has been growing interest in the effects of excited-state delocalizations on charge photogeneration processes~\cite{Kaake2012,Bakulin2012,Chen2013,Tamura2013a,Tamura2013b,Kaake2014,Kaake2015}.  
Therefore, we studied the excited-state dynamics in DNTT as a model system to provide insight into the role of the CT states in OSCs.

In this Letter, we present a theoretical study of optical properties and dissipative excitonic dynamics of DNTT. 
The optical absorption spectra and associated CT character are computed by introducing a tight-binding Hamiltonian that treats Frenkel and CT states on the same footing.
The excitation dynamics of the DNTT thin films was simulated by a stochastic Schr\"{o}dinger equation with spectral densities derived from molecular dynamics simulations and excited state quantum chemistry calculations.
Our simulations show coherent dynamics of the Frenkel and CT excitons in 50 fs after optical excitation.
The importance of charge delocalization in the mixing of CT states will be discussed.

\begin{figure*}[!th]
\begin{center}
 	\includegraphics[width=14cm]{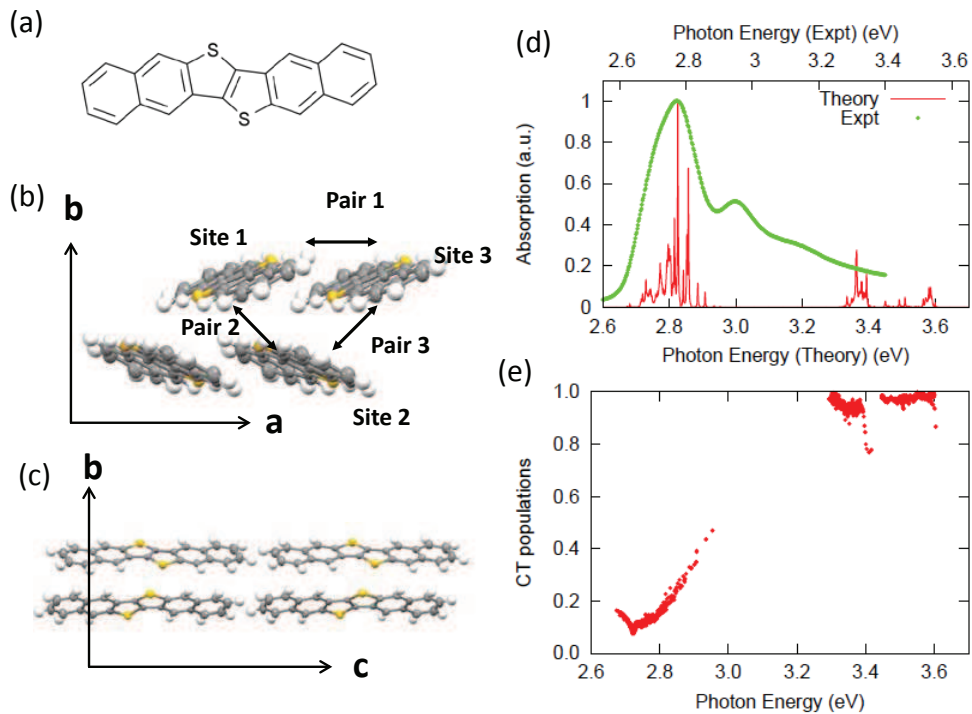}
\end{center}
\caption{(a) Chemical structure of DNTT. Crystal structure of DNTT in (b) $ab$ and (c) $bc$ planes. The crystal parameters are as follows: a = 6.187, b = 7.662, c=16.62 \AA; $\beta$=92.49$^\circ$~\cite{Yamamoto2007}. (d) Computed absorption spectrum of DNTT crystal in comparison with the experiment~\cite{Ishino2014}. (e) The CT populations of excited states.}
\end{figure*}

We adapt a tight-binding model of a system comprising of one electron and one hole in a molecular aggregate~\cite{Meier2007,Abramavicius2010,Kuhn2011,Saikin2013}. Each molecule has two frontier orbitals: the highest occupied molecular orbital (HOMO) and the lowest unoccupied molecular orbital (LUMO). A diabatic state of one electron and one hole, with localized wavefunctions, is obtained from creation operators as follows: $\bigl|e_ih_j \bigr>=\frac{1}{\sqrt{2}}\left(c^{\dagger}_{i\alpha}d^{\dagger}_{j\beta} + c^{\dagger}_{i\beta}d^{\dagger}_{j\alpha}\right)\bigl|0\bigr>$, where $\bigl|0\bigr>$ is an electronic ground state of the molecular aggregate, the operators $c^{\dagger}_{i\alpha}$ and $d^{\dagger}_{j\beta}$ create an electron of $\alpha$ spin on $i$th molecule (site) and a hole of $\beta$ spin on $j$th molecule, respectively. $\bigl|e_ih_i\bigr>$ represents a localized Frenkel state on $i$th molecule, while $\bigl|e_ih_j\bigr>$ denotes a CT state of an electron and a hole being localized on $i$th and $j$th molecules, respectively.

The electronic Hamiltonian in this one-electron one-hole basis can be written as follows:

\begin{equation}
\bigr<e_ih_j \bigr|H_e\bigl|e_kh_l \bigr>=\delta_{jl}t_{ik}^{(e)}+\delta_{ik}t_{jl}^{(h)}+\delta_{kl}\delta_{ij}(1-\delta_{ik})W_{ik}^{(f)}-\delta_{ik}\delta_{jl}W_{ij}^{(c)}.
\label{eqn:H_elec.}
\end{equation}
$t_{ij}^{(e/h)}$ are composed of electron/hole transfer integrals for off-diagonal elements and electron affinities/ionization potentials for diagonals. $W_{ij}^{(f)}$ are excitonic and $W_{ij}^{(c)}$ are e--h Coulomb interactions.
To calculate the electronic Hamiltonian, we follow Fujimoto~\cite{Fujimoto2012,Fujimoto2013} and adapt that methodology to the fragment molecular orbital methods~\cite{Mochizuki2005,Fedorov2009,Tanaka2014}.
Representative state energies and electronic couplings are shown in Table I and II. The transfer integrals and excitonic interactions were calculated from fragment molecular orbital calculations at the Hartree-Fock/6-31G* level. The localized excitation energy, ionization potential, and electron affinity were obtained from density functional theory at the B3LYP/aug-cc-pvdz level. Further details of the electronic structure calculations are shown in the Appendix A.

\begin{table}[!b]
\caption{Electron and hole transfer integrals, and excitonic interactions in meV for molecular pairs in the DNTT crystal structure in the $ab$ plane (see Figure 1(b) for pair labeling).}
\begin{tabular}{crrr} \\ \hline
         & electron & ~~~~hole & exciton  \\ \hline
Pair 1 &    18&    111            &   2                 \\ 
Pair 2 &    65&    141               &   41                 \\ 
Pair 3 &    49&     48          &     22                \\ \hline
\end{tabular} \\

\caption{Frenkel and CT energies in eV (see Figure 1(b) for site labeling).}
\begin{tabular}{crrrrrrr} \\ \hline
State   &  $e_1h_1$  & $e_1h_2$ & $e_1h_3$ & $e_2h_1$ & $e_2h_3$ & $e_3h_1$ & $e_3h_2$ \\ \hline
Energy  & 2.85   & 3.49  &  3.80 & 3.53 & 3.48   & 3.77  &  3.51\\ \hline
\end{tabular} \\
\end{table}

By diagonalizing the electronic Hamiltonian, we get an adiabatic wave function, $H_e\bigl| \psi_I \bigr>=E_I\bigl| \psi_I \bigr>$, where $\bigl| \psi_I \bigr>=\sum_{i,j}C^{I}_{i,j}\bigl|e_ih_j\bigr>$. 
We obtained the absorption spectrum from the corresponding transition dipole moments, $\mu_I =\sum_{i}C^{I}_{i,i}\mu_i$, where $\mu_i$ denotes a transition dipole moment of the $i$th molecule.
We use a 7 $\times$ 7 $\times$ 7 supercell containing 686 molecules for the model structure for the DNTT crystal. 
Each energy level is broadened by a convolution of a Lorentz function of one meV.
We also calculated CT populations of the $I$th adiabatic wave function, $p^{CT}_I = \sum_{i \neq j}\bigl| \bigl<e_ih_j | \psi_I \bigr> \bigr|^2$, which quantifies the probability that the electron and hole are found on different molecules.

Figure 1(d) shows the computed optical absorption spectrum of the DNTT crystal in comparison with the measurement from ref\cite{Ishino2014}.
The absorption spectrum is composed of the main excitation with the peak position of 2.83 eV, and relatively weak excitation at around 3.3 and 3.6 eV. The lower energy bands consist of Frenkel states partially mixed with the CT states. The latter two weak bands have mostly CT character, the intensities of which are borrowed from the lower Frenkel energy bands. The CT character of the brightest state is 22\%, and that of largest Frenkel band is 47\%. The red edge of the main excitation has lower CT character due to the energy difference with the CT states.
Our model predicts that the low energy exciton band has 8 to 47\% of CT character.
The computed absorption spectrum is in reasonable agreement with the experiment, while the peak at around 2.94 eV cannot be reproduced. This may be ascribed to vibrational progressions~\cite{Ishino2014}. However, it lies outside the scope of this letter to incorporate the vibrational degrees of freedom into the absorption spectra.

To investigate the quantum dynamics of the Frenkel--CT excitons in the DNTT, we apply the stochastic Schr\"{o}dinger equation formulated by Zhao and coworkers~\cite{Zhong2013,Han2014,Han2015,Abramavicius2014}. In this formalism, time evolution of an excitonic wave function coupled to a phonon bath is obtained as follows:

\begin{align}
i\hbar \frac{\partial}{\partial t}\bigl| \psi(t) \rangle= & H_e\bigl| \psi(t) \rangle + \sum_nL_nu_n(t)\bigl| \psi(t) \rangle \notag \\
                & -i\sum_n\left[L_n\int_0^td\tau C_n^{(0)}(\tau)e^{-{\frac{iH_e\tau}{\hbar}}}L_ne^{{\frac{iH_e\tau}{\hbar}}}  \right]\bigl| \psi(t) \rangle,
\label{sse}
\end{align}
where $L_n = \bigl|n \rangle \langle n\bigr|$. Here effects of the phonon bath are incorporated through a stochastic force $u_n(t)$ and a zero-temperature bath correlation function $C_n^{(0)}(t)$. Both of them are calculated from a spectral density $j_n(\omega)$ for the $n$th diabatic state~\cite{Zhong2013,Abramavicius2014}. The spectral densities~\cite{Damjanovic2002,Valleau2012} have been obtained by combing molecular dynamics simulations for the DNTT crystal with excited-state calculations using time-dependent density functional theory at B3LYP/6-31G* level. The details for calculating spectral densities are presented in the Appendix B.

An ensemble average of stochastic realization of eq (2) yields the two-body electron--hole density matrix,
\begin{equation}
\rho^{eh}(t) = \left<  \bigl| \psi(t) \rangle\langle \psi(t) \bigr| \right>_{ens}.
\end{equation}
A system of 18 DNTT molecules in the $ab$ plane was taken as a model of thin films. As illustrative examples, localized Frenkel ($\bigl|e_1h_1\bigr>$) and CT ($\bigl|e_1h_2\bigr>$) states are used for initial conditions for the simulations of dissipative quantum dynamics. See Figure 2(a) and (b) for the structure and site labeling. The state energies of the Frenkel and CT states are 2.85 and 3.49 eV, respectively. The Runge-Kutta method was used for numerical propagation with a time of 0.1 fs. The excitonic density matrix and physical properties are averaged over 5,000 trajectories at the temperature of 300 K.

To see dissipative dynamics of the 18 DNTT molecular system, Frenkel and CT populations, and electronic energies defined by $\langle \psi(t) \bigr|H_e\bigl| \psi(t) \rangle$ are shown in Figure 2(c) and (d). Thermal equilibration is achieved within 1.5 and 2.7 ps from the initial Frenkel and CT states, respectively, as they are converged to the similar electronic energy of 2.75 eV. From the initial CT state, the CT populations follow almost the same dynamics as the electronic energies; these plots reflect the relaxation dynamics from the higher CT state toward lower Frenkel states. On the other hand, from the initial Frenkel state, the CT populations is first increased and then decreased toward the equilibration value of around 0.1. 
As shown in Figure 2(d), the CT states are strongly mixed with the Frenkel states in the initial 50 fs---a mixed Frenkel--CT exciton is formed.

\begin{figure*}[!th]
\begin{center}
 	\includegraphics[width=11cm]{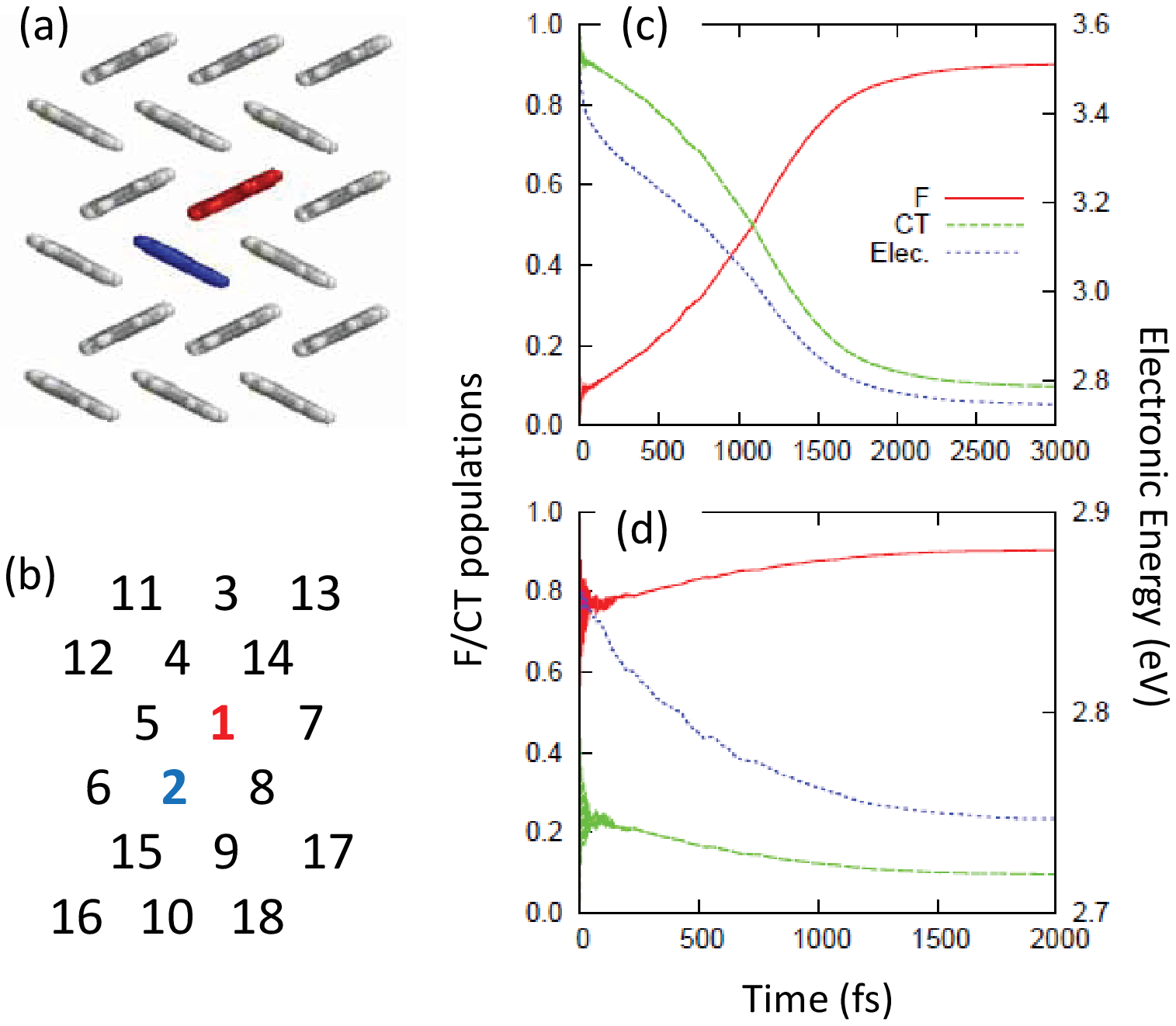}
\caption{(a) A structure and (b) site indices of 18 DNTT. Relaxation dynamics from (c) CT ($\psi(t=0)=\bigl|e_1h_2\bigr>$) and (d) Frenkel ($\psi(t=0)=\bigl|e_1h_1\bigr>$) states. In (c) and (d), Frenkel and CT populations and electronic energies are shown.}

 	\includegraphics[width=12cm]{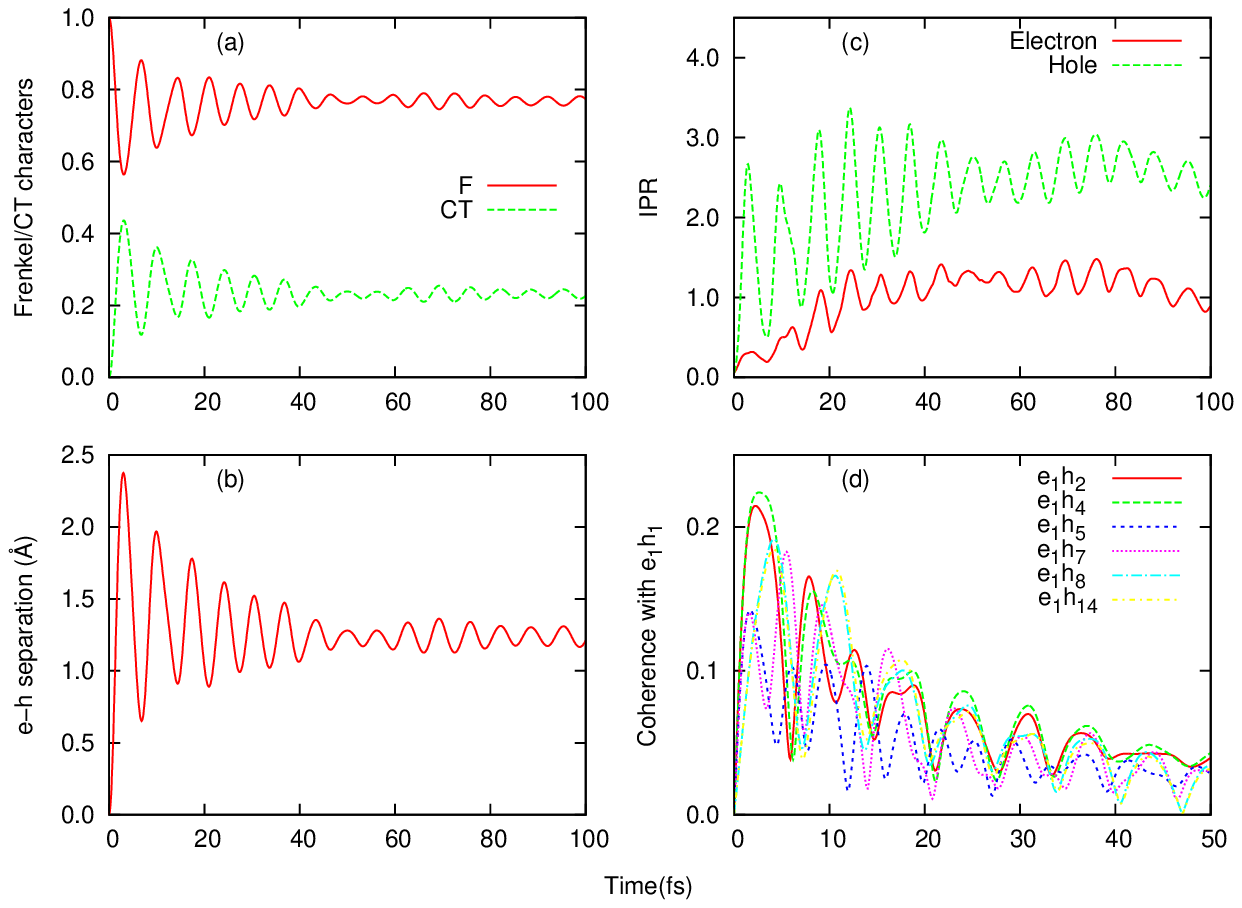}
\caption{Ultrafast dynamics from Frenkel state ($\bigl|e_1h_1\bigr>$): (a) Frenkel and CT populations. (b) Electron--hole separation (c) inverse participation ratio of electron and hole density matrix, and (d) absolute values of coherence with the initial Frenkel state. The time range of (d) is different from others.}
\end{center}
\end{figure*}

In what follows, we investigate the ultrafast time scale mixing of the CT states.
In addition to the Frenkel and CT populations, e--h separation, inverse participation ratios (IPRs) of the electron and hole density matrices, and coherence with the initial Frenkel state are shown in Figure 3 (a)--(d).  
The e--h separation of $\bigl|e_ih_j \bigr>$ is defined as the center of mass distance between $i$th and $j$th molecules.
To calculate the IPRs, one-body electron or hole density matrices were first calculated by tracing out the hole or electron basis:
$\rho^{e}(t) = Tr_h\rho^{eh}(t)$ or $\rho^{h}(t) = Tr_h\rho^{eh}(t)$. The IPRs of electron or hole density matrix is given by
\begin{equation}
L^{e/h}_{\rho}(t)= \frac{\left( \sum_{mn}\left| \rho^{e/h}_{mn}\right| \right)^2}{N\sum_{mn}\left| \rho^{e/h}_{mn}\right|^2}.
\end{equation}
These measures describe the delocalization of the electron or hole density matrix~\cite{Mukamel2000,Dahlbom2001}.

In Figure 3(a), the CT populations are increased and show oscillatory behavior which implies coherence among the Frenkel and CT states. The corresponding picture are also observed from the e--h separation: The e--h separation is first increased and subsequently decreased to the equilibrated value of around 0.5 \AA.
Because the hole transfer integrals are higher than those for the electrons, there is an increase in the e--h separation which is subsequently moderated by e--h Coulomb attractions.
This competition between the higher hole transfer and e--h Coulomb attraction results in the oscillation.

The effect of higher hole transfer is also seen by comparing electron and hole IPRs as represented in Figure 3(c), indicating that the hole IPR is greater than that of electron. 
The delocalization of the hole wave function is also confirmed by coherence among diabatic states, $\bigl|e_1h_i \bigr>$, as presented in Figure 3(d).
This delocalization can effectively weaken the e--h Coulomb attraction and thus become an initial driving force for increasing e--h separation and the mixing of CT states.
Localization of wave functions due to the nuclear vibrations works toward decreasing the e--h separations. 
Our finding is in line with the observations by Tamura and Burghardt~\cite{Tamura2013a,Tamura2013b} that charge separation at the donor/acceptor interface can be enhanced by charge delocalization. 


In the time-resolved spectroscopy experiment by Ishino et al.~\cite{Ishino2014}, the pump pulses excite to exciton manifolds at 3.1 eV, which leads to the optical excitation of the blue edge of the absorption maximum. The subsequent relxation to the lowest exciton state occurs within about 2.1 ps. Our simulations shows that the relaxation time from the localized Frenkel state, which is resonant to the absorption maximum, to the lowest exciton state is about 1.5 ps.  The relaxation time of 1.5 ps would qualitatively correspond to that observed by the experiments. However, simulating more extended systems would be necessary for quantitative comparison between the simulation and experiments.

The mixing of CT states in optical excitations of the DNTT has been proposed based on a derivative-like feature of transient absorption spectra~\cite{Ishino2014}:  Upon the excitation pulse of 3.1 eV the transition absorption spectra exhibit derivative-like features which are similar to Stark spectra. This result suggests the existence of transient charged species exerting electric fields on surrounding molecules. 
It follows that the CT excitons are formed after photoexcitation, the local electric fields of which induce Stark shifts of surrounding molecules. 
Relating our simulations with the experiments, we argue that the derivative-like features would appear after Frenkel--CT decoherence. 
If the optically-excited state is a superposition of Frenkel and CT states, the absorption of them cannot be disentangled. In other words, Stark effects from CT states occur after the Frenkel-CT decoherence time.
Our simulations predict that the derivative-like feature may appear about 50 fs after the optical excitation. A more rigorous comparison to the experimental observations has to be done by directly calculating the time-resolved spectroscopy signals~\cite{Mukamel1999}, which will be studied elsewhere.

We have presented a theoretical study on the optical properties and excitation dynamics of the DNTT thin films. The tight-binding Hamiltonian combined with ab initio calculations reasonably reproduces the experimental absorption spectrum and predicts that the low-energy Frenkel exciton band consist of 8 to 47\% CT character. The quantum dynamics simulations show the coherent dynamics of Frenkel and CT states about 50 fs after photoexcitation.
The oscillation and lifetime of the e--h separations are also analyzed in terms of charge delocalization/localization and the e--h Coulomb attractions.
Combined with large-scale quantum dynamics simulations~\cite{Huh2014,Sawaya2015}, the present approach can be applied to more extended systems such as organic bulk heterojunction solar cells. Such a device-level simulation will be useful to design and improve novel organic electronic materials.

\section*{ACKNOWLEDGMENT}
The authors thank Dr Joonsuk Huh for helpful discussion and implementation of stochastic Schr\"{o}dinger equations and Thomas Markovich for the calculation of spectral densities. T.F thanks Prof. Yuji Mochizuki for discussion on the electronic-structure calculations and Prof. Kazuya Watanabe and Mr. Shunsuke Tanaka for useful discussions on their experiments. The majority of our computation was carried out on Harvard University's Odyssey cluster, supported by the Research Computing Group of the FAS Division of Science. 
T.F acknowledge support from JSPS Grant-in-Aid for Scientific Research on Innovative Areas "Dynamical ordering of biomolecular systems for creation of integrated functions" (Grant No. 25102002).
T. F., N. S. and A. A.-G. acknowledge support from the Center for Excitonics, an Energy Frontier Research Center funded by the US Department of Energy, Office of Science and Office of Basic Energy Sciences under award DE-SC0001088.

\appendix

\section{Electronic structure calculations}
The electronic Hamiltonian in one-electron one-hole basis is given as follows:
\begin{equation}
\bigr<e_ih_j \bigr|H_e\bigl|e_kh_l \bigr>=\delta_{jl}t_{ik}^{(e)}+\delta_{ik}t_{jl}^{(h)}+\delta_{kl}\delta_{ij}(1-\delta_{ik})W_{ik}^{(f)}-\delta_{ik}\delta_{jl}W_{ij}^{(c)}.
\label{eqn:H_elec.}
\end{equation}
The parameters for this electronic Hamiltonian were obtained based on the method by Fujimoto~\cite{Fujimoto2012,Fujimoto2013}.
State energies of localized Frenkel and CT states can be calculated as follows:

\begin{equation}
\bigr<e_ih_i \bigr|H_e\bigl|e_ih_i \bigr> =  E_i^{*},
\end{equation}
\begin{equation}
\bigr<e_ih_j \bigr|H_e\bigl|e_ih_j \bigr> = -E_{i}^{EA} + E_{j}^{IP} - W_{ij}^{(c)}.                                                 
\end{equation}
Here, $E_i^{*}$ is a local excitation energy, $E_{i}^{EA}$ is an electron affinity, and $E_{i}^{IP}$ is an ionization potential.
The electron affinities and ionization potentials are defined as orbital energies or energy difference between neutral and charged molecules. 

Electronic interactions among Frenkel states are described as Coulomb interactions between transition densities of localized excitations of $i$th and $j$th molecules,
\begin{equation}
\bigr<e_ih_i \bigr|H_e\bigl|e_jh_j \bigr>  =  W_{ij}^{(f)}.
\end{equation}                                              
Those among Frenkel and CT states are transfer integrals,
\begin{equation}
\bigr<e_ih_i \bigr|H_e\bigl|e_jh_k \bigr> = b_{i}^{H \to L}\delta_{ij}t_{ik}^{(h)} + b_{i}^{H \to L}\delta_{ik}t_{ij}^{(e)}.
\end{equation}
In addition to the transfer integrals describing single electron or hole transfer steps, we adopt the HOMO-LUMO amplitude of the excited state, $b_{i}^{H \to L}$, to take account of the sequential process of the localized excitation of $i$th molecule and subsequent electron (hole) transfer from $i$th to $j$th molecules~\cite{Fujimoto2012}.

We consider single electron or hole transfer steps for interactions among CT states:
\begin{equation}
\bigr<e_ih_j \bigr|H_e\bigl|e_kh_l \bigr>=\delta_{jl}t_{ik}^{(e)}+\delta_{ik}t_{jl}^{(h)}.
\end{equation}
This term is nonzero unless $i=k$ or $j=l$. 

Due to the symmetry of the molecular crystal, we adopt same values for $E_{i}^{*}$, $E_{i}^{EA}$, and $E_{i}^{IP}$. The transfer integrals were considered for nearest-neighbor pairs as they are short-range interactions. The long-range Coulomb interactions, $W_{ij}^{(f)}$ and $W_{ij}^{(c)}$, are approximated as sum of pairwise interactions of atomic point charges,

\begin{equation}
W_{ij}^{(f)} = \sum_{A \in i}\sum_{B \in j}\frac{q_{A}^{(f)}q_{B}^{(f)}}{|\mathbf{R_A}-\mathbf{R_B}|},
\end{equation}                                              

\begin{equation}
W_{ij}^{(c)} = \sum_{A \in i}\sum_{B \in j}\frac{q_{A}^{(e)}q_{B}^{(h)}}{|\mathbf{R_A}-\mathbf{R_B}|},
\end{equation}         
where $\mathbf{R}$ denotes a nuclear position, and $q_{A}^{(f)}$ are transition atomic charges\cite{Madjet2006}, $q_{A}^{(e)}$ are electron atomic charges, and $q_{A}^{(h)}$ are hole atomic charges. Those atomic point charges were determined in such a way that they reproduce electrostatic potentials of transition ($q_{A}^{(f)}$), LUMO ($q_{A}^{(e)}$), or HOMO densities ($q_{A}^{(h)}$).

$E_{i}^{*}$, $E_{i}^{EA}$, and $E_{i}^{IP}$ were calculated at the B3LYP/aug-cc-pvdz level embedded in the presence of external point charges of surrounding molecules. These single-molecular properties are obtained by using Q-chem~\cite{QCHEM4}. The atomic charges describing the electrostatic potential of a DNTT molecule were calculated by the RESP method at the Hartree-Fock/6-31G* level. Transition, electron, and hole charges are obtained by a multi-layer fragment molecular orbital method at the configuration interaction singles/6-31G* level~\cite{Mochizuki2005}. The transfer integrals were calculated by the projective method developed by Kirkpatrick~\cite{Kirkpatrick2008,Baumeier2010}. Fragment molecular orbital calculations were performed by using ABINIT-MP~\cite{Tanaka2014}.

\section{Spectral densities}
Spectral densities (SDs) describe the frequency-dependent coupling strength between electron and phonon~\cite{May2011}. 
The coupling strength of CT states with a phonon bath would depend on the electron--hole distance: In small e--h separation, electron and hole interact together with a phonon bath as a Frenkel state. On the other hand, in large e--h separation, the electron and hole interact with a phonon bath individually. We need to treat $N$ state-specific SDs for CT states, which is computationally expensive to calculate all of them from mixed quantum/classical simulations~\cite{Valleau2012}.
Instead, we define an e--h distance dependent SD, $j(R)$, from Frenkel ($j_f$), electron ($j_e$), and hole ($j_h$) SDs. The SD of a CT state, the e--h distance of which is $R$, is defined as $j(R)$. The asymptotic behavior of $j(R)$ is $j(R) = j_f$ for $R \to 0$ and $j(R) = j_e + j_h$ for $R \to \infty$. We interpolate these limits with one decay parameter $\alpha$, assuming the $1/R$ dependence as follows, 
\begin{equation}
j(R) = \left(1-\exp{\left(-\frac{\alpha}{R}\right)} \right)j_f + \exp{\left(-\frac{\alpha}{R}\right)}(j_e + j_h).
\end{equation}

The SDs are obtained as Fourier transform of autocorrelation functions~\cite{Valleau2012},
\begin{equation}
j_{f/e/h} =  \left(\frac{\omega}{\pi k_BT}\right)\int_0^{\infty}\left<\Delta E(t)\Delta E(0) \right>cos(\omega t)dt,
\end{equation}
where $E=E^{*}$ for Frenkel, $E=E^{EA}$ for electron, and $E=E^{IP}$ for hole SDs.

A 4 x 4 x 3 supercell was first created based on the experimentally determined DNTT unit cell~\cite{Yamamoto2007}. Molecular dynamics simulations were carried out with a time step of 2 fs in the NVT ensemble for $T$ = 298 K with periodic boundary conditions. Other simulation conditions were same as that of ref~\cite{Sule2010} except that the non-bonded interactions were cut off at 12 \AA. After an equilibrium run of 5 ps, snapshots were collected at every 2 fs time steps along a production run of 10 ps. TDDFT B3LYP/6-31G* calculations were performed for those snapshots to obtain $E^{*}$, $E^{IP}$ and $E^{EA}$.  $E^{IP}$ and $E^{EA}$ were approximated as the HOMO and LUMO Kohn-Sham orbital energies within Koopmans' theorem. We adopt a superresolution method to extract SDs from the autocorrelation functions~\cite{Markovich2013}. Electron and hole SDs were scaled such that the reorganization energies of electron and hole became 101 and 65 meV~\cite{Sule2010}, respectively. The calculated SDs for the Frenkel state and charge-separated state ($j_e + j_h$) are shown in FIG. 4. The reorganization energy of the Frenkel SD was calculated to be 53 meV. The SDs of CT states are defined with $\alpha$ being 5 \AA, which is close to the center of mass distance between nearest-neighbor molecules. 

\vspace{1cm}
\begin{figure}[!th]
 	\includegraphics[width=6cm]{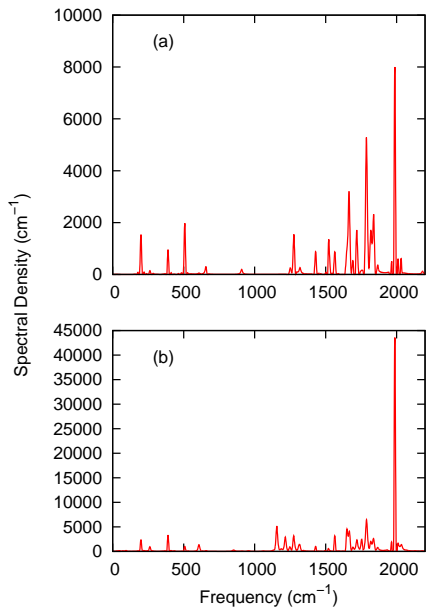}
\caption{SDs for (a) Frenkel state ($j_f$)and (b) charge-separated state ($j_e + j_h$).}
\end{figure}

\vspace{18cm}


\end{document}